\documentclass[a4paper]{article} \usepackage{amstex}
\newcommand{\cz}{{\mathbb C}}

\newcommand{\rz}{{\mathbb R}}

\def\tr{\mathop{\mathrm{tr}} \nolimits} 
\def\const{{\mathrm{const}}} 
\def\qed{\hbox {\hskip 1pt \vrule width 4pt height 6pt depth 1.5pt
        \hskip 1pt}}
\def\grad{\mathop{\mathrm{grad}} \nolimits} 
\def\atp{\bigwedge\limits} 
\newtheorem{theorem}{Theorem}

\newtheorem{lemma}{Lemma}

\date{March 28, 1995}

\title{Proof of the Strong Scott Conjecture for Atomic and Molecular Cores}
\author{Alexei Iantchenko\\
        Oslo
        \and
        Elliott H. Lieb\thanks{Work partially supported by US National
        Science Foundation grant PHY90 19433 A04}\\
        Princeton
        \and
        Heinz Siedentop\thanks{Work partially supported by Norges
        forskningsr\aa d,
        grant 412.92/008}\\
        Oslo}

\begin{document}
\maketitle
\begin{abstract} The strong Scott conjecture about the electron
density at a distance $1/Z$ from an atomic nucleus of charge $Z$ and
its generalization for molecules are proved.  The density, suitably
scaled, converges to an explicit limiting density as $Z \rightarrow
\infty$.  Both a weak convergence and a pointwise convergence on
spheres holds.
\end{abstract}

\section{Introduction}\label{s1} A great deal is known about the
ground states of large atoms in the framework of the non-relativistic
Schr\"odinger equation, with fixed (i.e., infinity massive) nuclei.
The leading term, in powers of the nuclear charge $Z$, is given
exactly by Thomas-Fermi theory, as was proved by Lieb and Simon
\cite{LiebSimon1977}; see \cite{Lieb1981} for a review.  This leading
term in the energy is proportional to $Z^{7/3}$, with the
proportionality constant depending on the ratio of $N/Z$, which is
assumed to be held fixed as $Z \rightarrow \infty$.  Here, $N$ is the
electron number.  Neutrality, i.e., $N=Z$, is not required, even
though it is the case of primary physical interest.  The
characteristic length scale for the electron density (in the sense
that all the electrons can be found on this scale in the limit $Z
\rightarrow \infty$) is $Z^{-1/3}$.  The fact that the true
quantum-mechanical electron density, $\rho_d$, converges (after
suitable scaling) to the Thomas-Fermi density, $\rho^{TF}$, as $Z
\rightarrow
\infty$ with $N/Z$ fixed was proved in \cite{LiebSimon1977}.  The
chemical radius, which is another length altogether, is believed, but
not proved, to be order $Z^0$ as $Z \rightarrow \infty$.

The first correction to the $Z^{7/3}$ law is not, as was formerly
supposed, the $Z^{5/3}$ corrections arising from exchange and
correlation effects and kinetic energy corrections on the $Z^{-1/3}$
scale.  Instead it is $Z^2 = Z^{6/3}$ and arises from extreme quantum
mechanical effects on the innermost electrons, which are at a distance
$Z^{-1}$ from the nucleus.  Among these the most important are the $K$
shell electrons.  It was Scott
\cite{Scott1952} who pointed this out and he also gave a formula for
the correction term to the energy,
\begin{equation} E^{Scott} = \frac q8 Z^2,
\end{equation} where $q$ is the number of spin states per electron (of
course $q=2$ in nature).  It is noteworthy that $E^{Scott}$ does not
depend on the fixed ratio $N/Z$, provided $N/Z\neq0$.  This fact
agrees with the idea that $E^{Scott}$ arises from the innermost
electrons, whose energies, in leading order, are independent of the
presence of the electrons that are further from the nucleus.  The
truth of (1) (i.e., the statement that the energy is $E^{TF} +
E^{Scott} + o(Z^2)$ for fixed $N/Z\neq0$) was proved in
\cite{SiedentopWeikard1987O,SiedentopWeikard1989} (upper and lower
bound) and by Hughes \cite{Hughes1990} (lower bound) in the atomic
case and by Bach \cite{Bach1989} in the ionic case. (For different
proofs and extensions of this result see \cite{SiedentopWeikard1991},
Fefferman and Seco
\cite{FeffermanSeco1995,FeffermanSeco1992,FeffermanSeco1994T,FeffermanSeco1994Th,FeffermanSeco1994,FeffermanSeco1993},
and Ivrii and Sigal \cite{IvriiSigal1993}.)

The ``strong Scott conjecture'', which we prove here, was made later
in
\cite{Lieb1981}, Equation (5.34).  It concerns the electron density
$\rho_d$ at distances of order $Z^{-1}$ from the nucleus and states
that in limit $Z \rightarrow \infty$ a suitably scaled $\rho_d$
converges to the sum of the squares of {\it all} the hydrogenic bound
states.  This function is defined in Section~2 below and is
extensively analyzed in
\cite{HeilmannLieb1995}.  (Previously, an upper bound for $\rho_d$ at
the origin of the correct form, namely $O(Z^3)$, was derived in
\cite{Siedentop1994A,Siedentop1994B}.)  We prove this convergence in
several senses, one of which is a ``pointwise'' convergence on
spheres.  In fact we go further and show that the individual angular
momentum densities converge to the hydrogenic values, thereby giving a
somewhat more refined picture of the ground state.

Our main proof strategy is the usual one.  We add $\epsilon$ times a
one-body test potential to our Hamiltonian and then differentiate the
ground state energy with respect to $\epsilon$ at $\epsilon = 0$ in
order to find $\rho_d$.  To obtain pointwise convergence the test
potential is a radial delta-function.  To control the energy we rely,
in part, on the results and methods in
\cite{SiedentopWeikard1987O,SiedentopWeikard1989}.

In the following we shall state and prove our theorems for the neutral
case $N=Z$.  We do so to avoid the notational complexity and
additional discussion required for $N/Z \not= 1$.  It is
straightforward, however, to generalize our results to $N/Z \not= 1$.

In the next section precise definitions, as well as our main theorems
are given.  Section 3 contains two lemmata about the difference in
energies with and without the test potential.  Since there are
infinitely many hydrogenic bound states, we need these estimates in
order to be able to show that the sum of the derivatives (with respect
to $\epsilon$) equals the derivative of the sum.  The strong Scott
conjecture for atoms is proved in Section 4.  Section~5 contains the
obvious extension to molecules.  The Appendix contains a few needed
facts about ground state energies.

\section{Definitions and Main Results}\label{s1a} The Hamiltonian of
an atom of $N$ electrons with $q$ spin states each and a fixed nucleus
of charge $Z$ located at the origin is given by
\begin{equation} H_{N,Z} = \sum_{\nu=1}^N\left(-\Delta_\nu - {Z\over
|\frak{r_\nu}|}\right) + \sum_{1 \leq \mu < \nu \leq N} {1 \over
|{\frak r}_\mu-{\frak r}_\nu|} ,
\label1
\end{equation} in units in which $\hbar^2 /2m=1$ and $\vert e \vert =
1$.  It is self-adjoint in the Hilbert space ${\mathfrak
H}_N:=\atp_{\nu=1}^N\left(L^2({\mathbb R}^3) \otimes \cz^q\right)$,
i.e., antisymmetric functions of space and spin.  A general ground
state density matrix, denoted by $d$, can be written as
\begin{equation}\label{d} d=\sum_{\nu =1}^M
w_\nu\mid\psi_\nu\rangle\langle\psi_\nu\mid,
\end{equation} where the $\psi_\nu$ constitute an orthonormal basis
for the ground state eigenspace and where the $w_\nu$ are nonnegative
weights such that $\sum_{\nu =1}^M w_\nu =1 $.  It is well known that
the ground state can be degenerate, e.g., it is for the carbon atom.
The corresponding one-electron density is the diagonal part of the
one-electron density matrix and is, by definition,
\begin{equation}
\rho_d({\frak r}) = N \sum_{\nu=1}^M w_\nu \sum_{\sigma_1,...,\sigma_N
= 1}^q
\int_{{\mathbb R}^{3(N-1)}}
\left|\psi_\nu({\frak r},{\frak r}_2, ...,{\frak
r}_N;\sigma_1,...,\sigma_N)\right|^2 d{\frak r}_2...d{\frak
r}_N.\label3
\end{equation} The density $\rho_{l,d}$ of angular momentum $l$
electrons at radius $r$ from the nucleus is given in terms of the
normalized spherical harmonics $Y_{l,m} (\omega)$.
\begin{multline}
\rho_{l,d}(r) = N \sum^l_{m=-l} \sum_{\nu=1}^M w_\nu
\sum_{\sigma_1,...,\sigma_N = 1}^q
\int_{{\mathbb R}^{3(N-1)}} \\
\left\vert \int_{{\mathbb S}^2} \overline{Y_{lm} (\omega)} \psi_\nu
(r\omega,{\frak r}_2, ...,{\frak r}_N;\sigma_1,...,\sigma_N)d\omega
\right\vert^2 d{\frak r}_2...d{\frak r}_N\label{3l}
\end{multline} where we write ${\frak r} = r\omega$ and $d\omega$
denotes the usual unnormalized surface measure on the two dimensional
sphere ${\mathbb S}^2$ with $(4\pi)^{-1}\int_{{\mathbb S}^2} d \omega
= \int_{{\mathbb S}^2} \vert Y_{l,m} \vert^2 d
\omega = 1$.

Throughout the paper we will write $\varphi^{TF}_{Z}(r)$ for the
Thomas-Fermi potential of electron number $N=Z$ and nuclear charge
$Z$, i.e., $$\varphi^{TF}_{Z}(r) = Z/r - \vert {\frak r} \vert^{-1} *
\rho_{Z}^{TF},$$ where $\rho^{TF}_{Z}$ is the nonnegative minimizer of
the Thomas-Fermi functional
$$\int_{\rz^3}(\frac35(6\pi^2/q)^{2/3}\rho^{5/3} ({\frak r}) - \frac
Z{|{\frak r}|} \rho ({\frak r})) d{\frak r} + D(\rho,\rho)$$ under the
condition $\int \rho = N = Z$ and with
$$D(\rho,\rho):=\frac12\int_{\rz^6}{\overline{\rho({\frak
r})}\rho({\frak s})\over|{\frak r} - {\frak s}|}d{\frak r}d{\frak
s}.$$ Both $\varphi_{Z}$ and $\rho^{TF}_Z$ are spherically symmetric,
i.e., they depend only on $r = \vert {\frak r} \vert$.  There is a
scaling relation $\varphi^{TF}_{Z} (r) = Z^{4/3} \varphi^{TF}_1
(Z^{1/3} r)$, where $\varphi^{TF}_1$ is the Thomas-Fermi potential for
$Z=1$ and ``electron number'' equal to 1.  Similarly, $\rho^{TF}_{Z}
(r) = Z^2 \rho^{TF}_1 (Z^{1/3} r)$.  This scaling shows that the
``natural'' length in an atom is $Z^{-1/3}$.  Note that the
Thomas-Fermi functional has a unique minimizer.

The Scott conjecture, on the other hand, concerns the length scale
$Z^{-1}$, where we expect the density to be of order $Z^3$ instead of
$Z^2$.  In terms of the ``true'' density defined in (4), we now define
\begin{equation}
\rho_Z ({\frak r}) := Z^{-3} \rho_d ({\frak r}/Z).
\end{equation} Likewise, we define the angular momentum density
\begin{equation}
\rho_{l,Z} (r) := Z^{-3} \rho_{l,d} (r/Z).
\end{equation}

To formulate the strong Scott conjecture we consider the angular
momentum $l$ states of a hydrogen atom $(Z=1)$ with radial Hamiltonian
\begin{equation} h^H_l := - \frac{d^2}{d r^2} + \frac{l(l+1)}{r^2} -
\frac1r
\end{equation} with normalized eigenfunctions $\psi^H_{l,n}$ (that
vanish at 0 and $\infty$) corresponding to {\it negative} eigenvalues
$e_{l,n}^H$. (The superscript $H$ denotes ``Hydrogen'' and
distinguishes $h^H_l$ from other radial Hamiltonians to be considered
later.)  The normalization is $\int^\infty_0
\vert \psi_{n,l} (r) \vert^2 dr = 1$.  We define the corresponding
density in the channel $l$ to be
\begin{equation}\label{LHL}
\rho_l^H(r) := q(2l+1)\sum_{n=0}^\infty |\psi_{l,n}^H(r)|^2/(4\pi
r^2);
\end{equation} the total density is then
\begin{equation}\label{LH}
\rho^H(r) = \sum_{l=0}^\infty \rho_l^H(r).
\end{equation}

Although we shall not be interested in detailed properties of
$\rho^H$, we note the following proved in
\cite{HeilmannLieb1995}: The sum over $l$ and $n$ defining $\rho^H
(r)$ is pointwise convergent for all $r$.  It is monotone decreasing
and it decays asymptotically for large $r$ as $1/(6 \pi^2 r^{3/2})$.
This large $r$ asymptotics meshes nicely with the {\it small} $r$
behavior of ${1 \over q} \rho^{TF}_1 (r)$.

Note: In \cite{HeilmannLieb1995} the operator $h^H$ is defined using
atomic units $\hbar^2/m=1$, i.e., with $\frac12(-d^2/dr^2+l(l+1)/r^2)$
instead of $-d^2/dr^2 + l(l+1)/r^2$. Note also that we have included
the factor $q$ in the definition of $\rho$ which was not done in
\cite{HeilmannLieb1995}. Thus some care is needed in comparing
formulae there with formulae here.

Various notions for the convergence of the rescaled density are
possible.  Our precise statements are Theorems \ref{t0} and \ref{t1}
below and Theorem \ref{t2} in Section~\ref{s4}.
\begin{theorem}\label{t0} {\bf (Convergence in angular momentum
channels).}  Fix the angular momentum $l_0$.
\begin{enumerate}
\item\label{item1} For positive $r$ \begin{equation}\label O
\lim_{Z\rightarrow \infty} \rho_{l_0,Z}(r) = \rho^H_{l_0}(r)
\end{equation} (pointwise convergence).
\item \label{item2} Let $V$ be an integrable functions on the positive
real line. Then we have the weak convergence
\begin{equation}\label{12}
\lim_{Z\rightarrow\infty} \int_0^\infty rV(r)\rho_{l_0,Z}(r)dr =
\int_0^\infty rV(r)\rho^H_{l_0}(r)dr.
\end{equation}
\end{enumerate}
\end{theorem}

\begin{theorem}\label{t1} {\bf (Convergence of the total density).}
\begin{enumerate}
\item\label{item3} Let $W$ be a bounded (not necessarily constant)
function on the unit sphere. Then, as $Z\rightarrow \infty$,
\begin{equation}\label W \int_{{\mathbb
S}^2}W(\omega)\rho_Z(r\omega)d\omega \rightarrow \rho^H(r)
\int_{{\mathbb S}^2}W(\omega)d\omega \end{equation} (pointwise
convergence of spherical averages).
\item \label{item4} Let $V$ be a locally bounded, integrable function
on $\rz^3$. Then, as $Z \rightarrow \infty$,
\begin{equation}\label{14}
\int_{\rz^3} |\mathfrak{r}| V({\frak r})\rho_Z({\frak r})d{\frak r}
\rightarrow
\int_{\rz^3} |\mathfrak{r}| V({\frak r})\rho^H(|{\frak r}|)d{\frak r}.
\end{equation}
\end{enumerate}
\end{theorem}

Remarks:
\begin{enumerate}
\item It is not really necessary to take a sequence of ground state
        density matrices. We could take just a sequence of states,
        $d_{N,Z}$, that is an {\it approximate ground state} in the
        sense that $$ {\tr{(H_{N,Z} d_{N,Z})} - E_{N,Z} \over Z^2}
        \rightarrow 0$$ as $Z\rightarrow\infty$.  Here $E_{N,Z}$ is
        the bottom of the spectrum of $H_{N,Z}$.  It might not be an
        eigenvalue, and it certainly will not be one if $N/Z$ is
        larger than 2.
\item It is important to note that $W$ and $V$ in (\ref{W}) and
(\ref{14}) need not be spherically symmetric. It might appear that
only the spherical averages of $W$ and $V$ are relevant, but this
would miss the point. Theorem \ref{t1} says, that in the limit
$Z\rightarrow\infty$ there is no way to construct a ground state or
approximate ground state that is not spherically symmetric on a length
scale $Z^{-1}$. For example, in the case of carbon there are ground
states that are not spherically symmetric and for which replacing $W$
by its spherical average changes the left side of (\ref W).
\item A word about pointwise convergence.  The one-body density {\em
matrix} $\gamma ({\frak r}, {\frak r}^\prime)$, which is defined as in
(\ref3) but with $\vert \psi_\nu (\cdots ) \vert^2$ replaced by
$$\psi_\nu ({\frak r}, {\frak r}_2, \dots , {\frak r}_N; \sigma_1,
\dots , \sigma_N)
\overline{\psi ({\frak r}^\prime, {\frak r}_2, \dots , {\frak r}_N;
\sigma_1, \dots ,
\sigma_N),}$$ is easily seen to be in the Sobolev space $H^1 ({\mathbb
R}^3)$ when $\gamma$ is considered as a function of each variable
separately. The trace theorem in Sobolev spaces then implies that the
function of $\omega$ on the sphere ${\mathbb S}^2, \gamma
(r\omega,r'\omega^\prime)$ is in $L^q ({\mathbb S}^2)$ for all $q \leq
4$.  Thus, the integrals in (4) are well defined and $\rho_d (r\omega)
= \gamma (r\omega, r\omega)$ is in $L^2 ({\mathbb S}^2)$.  It is also
easy to see that $\sqrt{\rho_{l,Z} (r)}$ is in $H^1 (0, \infty)$ and
hence it is a continuous function of $r$.  Since $\rho_d ({\frak r})$
is in $L^2 ({\mathbb S}^2)$ the integrals in (\ref{O}) and (\ref{W})
are well defined when $W \in L^2 ({\mathbb S}^2)$. --- If $\rho$ and
$\gamma$ belong to a ground state of $H_{N,Z}$ with $N\leq Z$ then
they are even continuous functions in all variables. This follows from
the regularity theorem of Kato and Simon (Reed and Simon
\cite{ReedSimon1978}, Theorem XIII.39) and the uniform exponential
decay of ground state eigenfunctions. This decay is implied as
follows. By Zhislin's theorem the atomic Hamiltonian has infinitely
many eigenvalues below the essential spectrum and the ground state
eigenspace has finite dimension. This implies that the ground state
energy is always a discrete eigenvalue which, in turn, implies
exponential decay of the ground state eigenfunctions according to
Theorem XIII.42 of
\cite{ReedSimon1978}.
\end{enumerate}

\section{Eigenvalue Differences of Schr\"odinger Operators Perturbed
on the Scale $1/Z$ \label{s2}}

It is well known, and will be seen more explicitly in Section 4, that
the eigenvalues of $H_{N,Z}$ can be controlled to within an accuracy
of $o(Z^2)$ by considering a one-body Schr\"odinger operator with the
spherically symmetric potential given by Thomas-Fermi theory.  In the
angular momentum $l$ channel, this is
\begin{equation} h^{TF}_{l,Z} = - \frac{d^2}{d r^2} +
\frac{l(l+1)}{r^2} -
\varphi^{TF}_{Z} (r).
\end{equation} (We suppress the dependence on $N$ in $h^{TF}_{l,Z}$
since $N=Z$.)  Closely related to $h^{TF}_{l,Z}$ is the unscreened
hydrogenic Hamiltonian
\begin{equation} h^H_{l,Z} = - \frac{d^2}{d r^2} + \frac{l(l+1)}{r^2}
- \frac{Z}{r}.
\end{equation}

In this section we want to study how the spectra of these operators
are shifted by the addition of a perturbing potential of the form
$$\epsilon U_Z (r) = \epsilon Z^2 U(Z r)$$ where $\epsilon$ is a small
parameter and where $U$ is some fixed function.  In particular, $U$
will be a radial delta function, $U(r) =
\delta (r-a)$ for some $a > 0$.

Both cases, $h^{TF}$ and $h^H$, will be considered together and we
write
\begin{equation} h_{l,\epsilon, Z} = - \frac{d^2}{d r^2} + V_{l,Z} (r)
- \epsilon U_Z (r),
\end{equation} in which $V_{l,Z} = -Z/r + l(l+1)/r^2$ or $V_{l,Z} =
-\varphi^{TF}_{Z} (r) + l (l+1)/r^2$.

Our first lemma estimates the difference in the spectra of $h_{l,0,Z}$
and $h_{l, \epsilon, Z}$ by the difference in the trace ($\tr$) of the
negative parts $(h_{l, \epsilon, Z})_-$ or $(h_{l, 0, Z})_-$ (i.e.,
the sums of the negative eigenvalues).  This lemma will later on allow
us to interchange the limits $Z\rightarrow \infty$ and $\epsilon
\rightarrow 0$ with the $l$ summation.
\begin{lemma}\label{l0} Set $U(r)=\delta(r - a)$, $U_Z(r) = Z^2 U(Z
r)$ and assume $|\epsilon|\leq
\pi/(16a)$. Then $$|\tr(h_{l,0,Z})_- - \tr(h_{l,\epsilon,Z})_-| \leq
\epsilon
\frac{9a Z^2}{(l+1)^2(2l+1)}.$$
\end{lemma} Proof: By the minimax principle we have for $\epsilon >0$
\begin{equation}\label{differenz} 0\leq
s_{\epsilon,l,Z}:=\tr(h_{l,0,Z})_- - \tr(h_{l,\epsilon,Z})_- \leq
\epsilon\tr(U_Z d_{l,\epsilon,Z}).
\end{equation} Inserting the identity twice in the right side of
\ref{LH} we have
\begin{equation}\label{s_} s_{\epsilon,l,Z}\leq \epsilon \tr\left(ABC
B^* A^*\right) \leq \epsilon
\|A\|^2_\infty\|B\|^2_\infty \ \tr C
\end{equation} with
\begin{eqnarray*} A&:=&d_{l,\epsilon,Z}(h_{l,\epsilon,Z} +
c_{l,Z})^{1/2} \geq 0, \\ B&:=&(h_{l,\epsilon,Z} +
c_{l,Z})^{-1/2}(H_{0,l} + c_{l,Z})^{1/2},\\ C&:=&(H_{0,l} +
c_{l,Z})^{-1/2}U_Z(H_{0,l} + c_{l,Z})^{-1/2} \geq 0,
\end{eqnarray*} where $c_{l,Z}$ is any positive number bigger than
$|\inf \sigma(h_{l,
\epsilon,Z})|$, where $\sigma(h)$ denotes the spectrum of $h$. We also
define $H_{0,l}:=-d^2/d r^2 + l(l+1)/r^2$ to be the free operator in
the angular momentum channel $l$. Since $\varphi_{Z}^{TF} (r) \leq
Z/r$ and since $\inf \sigma (H_{0,l} - Z/r) = - Z^2/[4 (l+1)^2]$ we
can take $c_{l,Z}:= Z^2/(l+1)^2$ provided $\epsilon$ is not too large.

We now estimate these norms individually:

Because $c_{l,Z}$ is bigger than the modulus of the lowest spectral
point of $h_{l,\epsilon,Z}$ and $d_{l,\epsilon,Z}$ is the projection
onto the negative spectral subspace of $h_{l,\epsilon,Z}$ we have
\begin{equation}\label A
\|A\|_\infty \leq \sqrt{c_{l,Z}}.
\end{equation}

For $B$ we get
\begin{multline}\label{B}
\|B\phi\|^2 = \|(h_{l,\epsilon,Z} + c_{l,Z})^{-1/2}(H_{0,l} +
c_{l,Z})^{1/2}\phi\|^2\\ = (\phi,(H_{0,l} +
c_{l,Z})^{1/2}(h_{l,\epsilon,Z} + c_{l,Z})^{-1}(H_{0,l} +
c_{l,Z})^{1/2}\phi)\\ = (\phi,{1 \over 1 - W_{l,\epsilon,Z}}\phi)
\end{multline} with $$W_{l,\epsilon,Z}:=(H_{0,l} +
c_{l,Z})^{-1/2}(\varphi_Z + \epsilon U_Z)(H_{0,l} + c_{l,Z})^{-1/2}.$$
We will then have
\begin{equation}\label{bpunch}
\|B\| \leq \sqrt2
\end{equation} if we can show that $W_{l, \epsilon, Z}$ is bounded
above by $\frac12$. To this end we note that $H_{0,l}+c_{l,Z}$ is
invertible, so that we can write any normalized $\phi\in L^2(\rz^+)$
as $\phi:= (H_{0,l} + c_{l,Z})^{1/2} \psi/\|(H_{0,l} + c_{l,Z})^{1/2}
\psi\|$ with $\psi$ in the domain of $H_{0,l}$.  Thus, we have to show
that $$ (\phi,W_{l,\epsilon,Z}\phi) = (\psi,(\varphi_Z + \epsilon
U_Z)\psi)/(\psi, (H_{0,l} + c_{l,Z})\psi) \leq \frac12,$$ which is
equivalent to
\begin{equation}\label{z}
\frac12 (H_{0,l} + c_{l,Z}) - \varphi_Z - \epsilon U_Z \geq 0.
\end{equation}

Since $\varphi^{TF}_{Z} (r) \leq Z/r$ and since
\begin{multline}\label{NN} (\psi,U_Z\psi) = Z|\psi(a/Z)|^2 \leq Z
\int_0^{a/Z}{|\psi|^2}'(r) d r
\leq 2Z \Re\int_0^{a/Z}\psi(r)\psi'(r) d r \\
\leq 2Z \|\psi'\|_2 \left\{ \int^{a/Z}_0 \psi (r)^2 d r \right\}^{1/2}
\end{multline} we have that $(\psi, U_Z \psi) \leq (4a/\pi) \Vert
\psi^\prime \Vert^2_2$.  (Here we use the inequality, $\int^L_0
\psi^{\prime 2} \geq (\pi /2L)^2
\int^L_0 \psi^2$, when $\psi (0) = 0$.)  Thus, (\ref z) is implied by
\begin{equation}\label y -4\frac{Z^2}{4(l+1)^2} + c_{l,Z} + \inf
  \sigma ((\frac14-\frac{4\epsilon a}\pi)H_{0,l}) \geq 0 .
\end{equation} The sum of the first two terms in (\ref y) vanishes
because of our choice of $c_{l,Z}$ and last term in (\ref y) is zero
when $\epsilon \leq \pi/(16a)$.  This is true by hypothesis, and the
bound on the norm of $B$ is proved.

Finally the trace of $C$ is computed easily, since it is of rank one.
Since the kernel $(H_{0,l}+c_{l,Z})^{-1} (r,r')$, is a positive,
continuous function in both variables, $\tr C = Z (H_0 + c_{l,Z})^{-1}
(a/Z, a/Z)$.  A well known calculation yields
$$(H_{0,l}+c_{l,Z})^{-1}(r,r') =\sqrt r K_{l+\frac12}(\sqrt{c_{l,Z}} \
r_>) I_{l+\frac12}(\sqrt{c_{l,Z}} \ r_<)\sqrt{r'},$$ where $r_> =
\max\{r,r'\}$ and $r_< = \min\{r,r'\}$. Thus $$\tr C = a
K_{l+\frac12}(\sqrt{c_{l,Z}} \frac{a}{Z}) I_{l+\frac12}(\sqrt{c_{l,Z}}
\frac{a}{Z}).$$ The modified Bessel functions $I_{l+\frac12}$ and
$K_{l+\frac12}$ are both positive and the following uniform asymptotic
expansions hold.  (See Olver \cite{Olver1961} for a proof of the
estimates of the remainder terms, \cite{Olver1962}, p. 6 for the
remainder in the form used here,
\cite{Olver1974}, Chapter 10, Paragraph 7 for a review; see also Olver
\cite{Olver1968}, section 9.7.)
\begin{eqnarray} K_{n}(n x) &=& \sqrt{\pi t \over 2n} e^{-n\xi}
\left[1 +
\epsilon_{0,2}(n,t)\right]\label K\\ I_{n}(n x) &=&
\sqrt{\frac{t}{2\pi n}}{e^{n\xi} \over 1 - \epsilon_{0,1}
(n,0)}\left[1 + \epsilon_{0,1}(n,t)\right],\label I
\end{eqnarray} where
\begin{eqnarray*}
\xi&:=& \frac1t -\frac12 \log\frac{1+t}{1-t}\\ t &:=&
(1+x^2)^{-\frac12}
\end{eqnarray*} and
\begin{eqnarray*} |\epsilon_{0,1}(n,t)| &\leq& \frac{n_0}{n-n_0}\\
|\epsilon_{0,2}(n,t)| &\leq& \frac{n_0}{n-n_0}
\end{eqnarray*} with $n_0 := \frac1{6\sqrt5} + \frac1{12} \leq 1/6$.
Thus, $$K_n(n x)I_n(n x) \leq
\frac1{2n(1+x^2)^{\frac12}}\frac{n^2}{(n-2n_0) (n-n_0)}\leq
\frac9{4n},$$ where the last inequality holds for $n\geq\frac12$.
Thus
\begin{equation}\label{cpunch}
\tr C \leq \frac92\frac {a}{(2l+1)}.
\end{equation}

Putting (\ref{s_}),(\ref A), (\ref{bpunch}), and (\ref{cpunch})
together yields $$s_{\epsilon,l,Z} \leq \epsilon
\frac{9c_{l,Z}a}{(2l+1)}\leq
\epsilon\frac{9a Z^2}{(l+1)^2(2l+1)}$$ which is more than the desired
result for $\epsilon > 0$.

If $\epsilon$ is negative we have, again by the minimax principle,
\begin{equation}\label{differenzn} 0\geq
s_{\epsilon,l,Z}:=\tr(h_{l,0,Z})_- - \tr(h_{l,\epsilon,Z})_- \geq
\epsilon\tr(U_Z d_{l,0,Z}).
\end{equation} Similar to the previous analysis, we have
\begin{equation}\label{s_n} s_{\epsilon,l,Z} \geq \epsilon \tr\left(D
C D^*\right) \geq \epsilon
\|D\|^2_\infty \tr C
\end{equation} with $$ D:=d_{l,0,Z}(h_{l,0,Z} + c_{l,Z})^{\frac12}.$$
As for $A$ above, we have $$\|D\|_\infty \leq \sqrt{c_{l,Z}}.$$
Putting this together with (\ref{cpunch}) gives $$s_{\epsilon,l,Z}
\geq \epsilon \frac{9c_{l,Z}a}{2(2l+1)}\geq
\epsilon\frac{9a Z^2}{2(l+1)^2(2l+1)}$$ which is even better than the
desired result for negative $\epsilon$.
\qed

The next result will later on allow us to interchange the limits
$Z\rightarrow \infty$ and $\epsilon \rightarrow 0$ with the $n$
summation for fixed $l$.
\begin{lemma}\label{l1} Set $U(r)=\delta(r - a)$ and assume
$|\epsilon|\leq \pi/(4a)$, $a>0$. Let
\begin{equation}\label{HH} h_{l, \epsilon} := -{d^2\over d r^2} +
{l(l+1) \over r^2} - \frac{1}{r} -\epsilon U (r)
\end{equation}
with form domain $H_0^1(0,\infty)$. Let $e_{n,l,\epsilon}$ denote the $n$-th
eigenvalue of
$h_{l,
\epsilon}$.  Then
\begin{equation}\label{HHa} |e_{n,l,0} - e_{n,l,\epsilon}|\leq
\frac1{(n+l)^2}\frac{\epsilon a}{\pi - 4\epsilon a}
\end{equation}
\end{lemma} Proof: For any $\psi$ in $H^1_0 (0, \infty)$ we have
$$|\psi(a)|^2 \leq 2 \frac2\pi a\|\psi'\|^2_2,$$ as proved in
(\ref{NN}) of Lemma 1.  Thus, for $\epsilon > 0$,
\begin{equation}\label{NNN} h_{l, \epsilon} \geq (1 - \frac{4\epsilon
a}\pi)\left[-{d^2\over d r^2} +
\frac{l(l+1)}{r^2} - \frac1{(1 - \frac{4\epsilon a}{\pi})r}\right].
\end{equation} This implies $$e_{n,l,\epsilon} \geq (1 -
\frac{4\epsilon a}{\pi}) \widetilde e_{n,l, 0}$$ where $\widetilde
e_{n, l, 0}$ is the $n$-th eigenvalue of [ \ \ ] in (\ref{NNN}), i.e.,
where the potential $r^{-1}$ is replaced by $(1-4 \epsilon a/\pi)^{-1}
r^{-1}$.  Thus, $$ 0\leq e_{n,l,0} - e_{n,l,\epsilon} \leq
\frac1{4(n+l)^2}
\left( -1 + (1- \frac{4\epsilon a}{\pi})^{-1}\right) =
\frac1{(n+l)^2}\frac{\epsilon a}{\pi - 4\epsilon a}, $$ which proves
the claim when $0 < \epsilon<\pi/(4a)$.

If $\epsilon$ is negative we have $$h_{l, \epsilon} \leq (1 -
\frac{4\epsilon a}{\pi})\left[-{d^2\over d r^2} + \frac{l(l+1)}{r^2} -
\frac1{(1 - \frac{4\epsilon a}{\pi})r}\right]$$ which again proves the
claim (by the same argument) when $0 > \epsilon>-\pi/(4a)$. \qed

\section{Proof of the Strong Scott Conjecture} \label{s3} We are now
able to give the proof of our theorems.  We begin with the proof of
Theorem \ref{t0} and begin with the first statement:

1. {\it The proof of the convergence of the spherical averages:} Set
$U (r) := \delta (r-a)$ for $a>0$ and $U_Z(r) := Z^2 U(Z r) = Z \delta
(r -
\frac{a}{Z})$.  Fix $l_0$ and let
\begin{equation}\label{34a} H_{N,Z}^\epsilon:=H_{N,Z} - \epsilon \sum
\limits^N_{\nu =1} U_Z(r_\nu)\Pi(l_0)
\end{equation} where $\Pi(l_0)$ denotes the projection onto angular
momentum $l_0$.  We define $\lambda(Z)$ -- which does not depend on
$\epsilon$ -- by
\begin{equation}\label{basis}
\lambda(Z):= a^2\rho_{l_0}(a) =
\frac{\tr(H_{N,Z}d) - \tr(H_{N,Z}^\epsilon d)}{\epsilon Z^2}.
\end{equation} Let us define $e^H_{n,l,\gamma,Z}$ and
$e_{n,l,\gamma,Z}$, $n=1,2,...$, $\epsilon \in \rz$, to be the
negative eigenvalues of the operators
\begin{eqnarray}\label{nackterop} H^H_{l,\epsilon,Z}&:=& -\frac{d^2}{d
r^2}+\frac{l(l+1)}{r^2}- \frac Z r-\epsilon U_Z \delta_{l,l_0}\\
H_{l,\epsilon,Z}&:=& -\frac{d^2}{d r^2}+\frac{l(l+1)}{r^2}-
\varphi_Z^{TF}-\epsilon U_Z\delta_{l,l_0}
\label{angezogenerop}
\end{eqnarray} with zero Dirichlet boundary on $(0,\infty)$.  To
obtain an upper bound for $\lambda(Z)$ we pick $\epsilon$ positive and
estimate as follows: by (\ref{Scott}) we have the upper bound
\begin{multline}\label{scottl}
\tr(H_{N,Z}d)\\
\leq \sum_{l=0}^{L-1}q(2l+1)\sum_n e^H_{n,l,0,Z} + \sum_{l=L}^\infty
q(2l+1)\sum_n e_{n,l,0,Z} -D(\rho_{TF},\rho_{TF})+ \const
Z^{\frac{47}{24}}
\end{multline} where $L=[Z^{1/9}]$.

To obtain a lower bound on $\tr (H^\epsilon_{N,Z}d)$ we first use the
lower bound \cite{LiebThirring1975,Lieb1981} on the correlations,
namely $- \const [N \int\rho^{5/3}_d]^{1/2}$, to reduce it to a radial
problem. Using the fact that $Z/r \geq \varphi^{TF}_{Z}(r)$ for $r >
0$ it follows from this that
\begin{multline}\label{scottla}
\tr(H_{N,Z}^\epsilon d)\\ \geq \sum_{l=0}^{L-1} q(2l+1)\sum_n
 e^H_{n,l,\epsilon,Z} + \sum_{l=L}^\infty q(2l+1) \sum_n
 e_{n,l,\epsilon,Z} - D(\rho_{TF},\rho_{TF}) -\const Z^{\frac53}.
\end{multline}
Note that (\ref{scottla}) arises from a relatively simple lower bound
calculation. Part of the proof of the Scott conjecture amounts to
proving that the right hand of (\ref{scottla}) is accurate to
$o(Z^2)$. This proof was carried out in \cite{SiedentopWeikard1989}
(see also \cite{Hughes1990,SiedentopWeikard1991}). We are {\it not}
rederiving the Scott correction for the energy, and it is not
necessary for us to do so here.

Define
$$ \theta(n) := \begin{cases} 1 & n>0\\
                              0 & n\leq0
                \end{cases}.$$
Since the eigenvalues of the perturbed problem ($\epsilon \neq 0$) are
equal to the unperturbed one ($\epsilon=0$) except for $l=l_0$, we get
the inequality
\begin{multline}
\limsup_{Z\rightarrow\infty}\lambda(Z)\\
\leq \liminf_{\epsilon\searrow0}\limsup_{Z\rightarrow\infty}
       \left[q(2l_0+1)\frac{\tr(H^H_{l_0,0,Z})_- - \tr(H^H_{l_0,\epsilon,Z})_-}
{\epsilon Z^2} \theta(L -l_0) \right.\\
\left.+\frac{\tr(H_{l_0,0,Z})_- -\tr(H_{l_0,\epsilon,Z})_- }{\epsilon Z^2}
\theta(l_0-L)
+ \const Z^{-\frac{1}{24}}\epsilon^{-1}\right]\label{o2}.
\end{multline}
Because $L$ eventually becomes larger than the fixed $l_0$,
\begin{multline}
\limsup_{Z\rightarrow\infty}\lambda(Z)
\leq
q(2l_0+1)\liminf_{\epsilon\searrow0}\limsup_{Z\rightarrow\infty}\frac{\tr(H^H_{l_0,0,Z})_- - \tr(H^H_{l_0,\epsilon,Z})_-}{\epsilon Z^2}\\
= q(2l_0+1)\liminf_{\epsilon\searrow0}
\frac{\tr(H^H_{l_0,0,1})_- - \tr(H^H_{l_0,\epsilon,1})_-}{\epsilon}\label{o4},
\end{multline}
where the last equation holds since because of the scaling of $h_{l_0,0,Z}^H$
and $h_{l_0,\epsilon,Z}$. Therefore
\begin{eqnarray}
&&\limsup_{Z\rightarrow\infty}\lambda(Z)\nonumber\\
&\leq& q(2l_0+1)\liminf_{\epsilon\searrow0} \sum_n\frac{e^H_{n,l_0,0,1} -
e^H_{n,l_0,\epsilon,1}}{\epsilon}\label{o5}\\
&=& q(2l_0+1)\sum_n\liminf_{\epsilon\searrow0} \frac{e^H_{n,l_0,0,1} -
e^H_{n,l_0,\epsilon,1}}
{\epsilon}\label{o6}\\
&=& a^2\rho^H_{l_0}(a).\label{o7}
\end{eqnarray}
To exchange the limit $\epsilon \searrow 0$ with the summation in
(\ref{o5}) we use Lemma \ref{l1}, which provides a summable majorant
for the series that is uniform in $\epsilon$ and thus allows us to
fulfill Weierstra\ss ' criterion for uniform convergence. Finally, to
deduce (\ref{o7}) from (\ref{o6}) we use the fact that the
one-dimensional delta potential is a relatively form bounded
perturbation, i.e., defines an analytic family in the sense of Kato.

To obtain a lower bound for $\lambda(Z)$ we pick $\epsilon$ negative
instead of positive and take the limit $\limsup_{\epsilon \nearrow 0}$
and $\liminf_{Z\rightarrow\infty}$ instead of
$\liminf_{\epsilon\searrow0}$ and
$\limsup_{Z\rightarrow\infty}$. Repeating the same steps gives the
same result except for reversing the inequalities, thereby yielding
the same bound (\ref{o7}) from below. This establishes the first claim
of the theorem.

2. {\it Proof of the weak convergence}: Because of the linear
dependence of the right and left hand side of (\ref{12}), it suffices
to prove the claim for the positive and negative parts of $V$
separately. Thus we may -- and shall -- assume that $V$ is positive.
We can now roll the proof back to the previous case as follows: First
we pick $Z$ large enough so that $l_0 < L$. It is convenient now, to
replace $\epsilon$ be $\epsilon/a$ in order that the right side of
(\ref{HHa}) in Lemma \ref{l1} is uniformly bounded in $a$ and
$\epsilon$ for all $a \in (0,\infty)$ and for $|\epsilon|\leq
\pi/8$. Then we integrate the inequality
\begin{equation}
a^2\rho_{l_0}(a) \leq
\frac{\tr(H^H_{l_0,0,Z})_- - \tr(H^H_{l_0,\epsilon/a,Z})_-}{(\epsilon/a)
Z^2}\theta(L -l) + \const/((\epsilon/a) Z^{\frac1{24}}),
\end{equation}
i.e.,
\begin{equation}\label{42a}
a\rho_{l_0}(a) \leq
\frac{\tr(H^H_{l_0,0,Z})_- - \tr(H^H_{l_0,\epsilon/a,Z})_-}{\epsilon
Z^2}\theta(L -l) + \const/(\epsilon Z^{\frac1{24}})
\end{equation}
against $V(a)$ from $0$ to $\infty$. Thanks to Lemma \ref{l0} the
right of side of (\ref{42a}) is bounded by $\const\ a$ and hence the
integral is finite. Next we write out the traces appearing in
(\ref{o7}) in terms of the eigenvalues and then use Lemma \ref{l1} to
provide a bound that is summable (over the eigenvalues) and integrable
(from $0$ to $\infty$), if $|\epsilon| < \pi/8$. This bound is
uniformly bounded in $\epsilon$, and so, by dominated convergence, we
can take the limit $\epsilon \searrow 0$ term by term. Using the
result (\ref O), which we established above, Equation (\ref{12}) is
now verified.  \qed

{\it Proof of Theorem \ref{t1}}: As was the case in the proof of
Theorem \ref{t1} we shall assume that $W$ and $V$ are nonnegative. For
Part 1 we proceed as for Theorem \ref{t1} and define
$H_{N,Z}^\epsilon$ as in (\ref{34a}), but with $\Pi(l_0)$ replaced by
$W(\omega)$. First we treat the case $W(\omega)=1$. We follow the
proof of Theorem \ref{t0} up to Equations (\ref{nackterop}) and
(\ref{angezogenerop}) (with $\delta_{l,l_0}$ replace by $W$). Then we
obtain analogously
\begin{eqnarray}\label{obenpunch}
&& \limsup_{Z\rightarrow\infty}\lambda(Z)\\
&\leq& \liminf_{\epsilon\searrow0}\limsup_{Z\rightarrow\infty}
       \left[\sum_{l=0}^\infty q(2l+1)
\frac{\tr(H^H_{l,0,Z})_- - \tr(H^H_{l,\epsilon,Z})_-}{\epsilon Z^2}
\theta(L -l) \right.\\
&& \left.+\sum_{l=0}^\infty q(2l+1)\frac{\tr(H_{l,0,Z})_-
-\tr(H_{l,\epsilon,Z})_- }{\epsilon Z^2} \theta(l-L)
+ \const Z^{-\frac{1}{24}}\epsilon^{-1}\right]\label{obenpunch2}\\
&=& \sum_{l=0}^\infty q(2l+1)
 \liminf_{\epsilon\searrow0} \limsup_{Z\rightarrow\infty}
\frac{\tr(H^H_{l,0,Z})_- - \tr(H^H_{l,\epsilon,Z})_-}{\epsilon
Z^2}\label{obenpunch3}\\
&=& \sum_{l=0}^\infty q(2l+1)
 \liminf_{\epsilon\searrow0}
\frac{\tr(H^H_{l,0,1})_- -
\tr(H^H_{l,\epsilon,1})_-}{\epsilon}\label{obenpunch4}\\
&=& a^2\sum_{l=0}^\infty \liminf_{\epsilon\searrow0}
\rho^H_l(a)\label{obenpunch5}\\
&=& a^2\rho^H(a).\label{obenpunch6}
\end{eqnarray}
To obtain (\ref{obenpunch2}) we use inequalities (\ref{scottl}) and
(\ref{scottla}). To obtain (\ref{obenpunch3}) we use the fact that
Lemma \ref{l0} provides a majorant uniform in $\epsilon$ and $Z$ which
is absolutely summable with respect to $\sum_{l=0}^\infty q(2l+1)$,
i.e., fulfills the Weierstra{\ss} criterion for uniform convergence
(or the hypothesis of Lebesgue's dominated convergence theorem), and
therefore allows the interchange of the limit and the $l$ summation,
and that the second sum tend term by term to zero. To obtain
(\ref{obenpunch4}) we use the fact that the eigenvalues of the bare
problem scale like $Z^2$. Finally, the convergence result of Theorem 2
was used to obtain (\ref{obenpunch5}).

To obtain a lower bound we pick $\epsilon$ negative instead of
positive and take the limit $\limsup_{\epsilon \nearrow 0}$ and
$\liminf_{Z\rightarrow\infty}$ instead of
$\liminf_{\epsilon\searrow0}$ and
$\limsup_{Z\rightarrow\infty}$. Repeating the same steps gives the
same result except for reversing the inequalities thereby yielding the
bound from below.

Let $W$ now be a general bounded, measurable function on the unit sphere
which we may -- according to the remarks in the beginning -- assume to be
positive. We take $\|W\|_\infty$=1.

Let us try to imitate the steps (\ref{obenpunch}) to
(\ref{obenpunch6}). As before we are faced with estimating the
eigenvalues of the one-body operators $H^H_{\epsilon,Z}:= -\Delta -
Z/|.| - \epsilon Z W\delta_{\frac aZ}$ and $H^{TF}_{\epsilon,Z}=
-\Delta - \varphi^{TF}_Z - \epsilon Z W\delta_{\frac aZ}$ but unlike
the previous case these cannot be simply indexed by the angular
momentum $l$ when $\epsilon \neq 0$; indeed the one-body operators
cannot be reduced to a direct sum of radial Schr\"odinger operators as
in (\ref{nackterop}) and (\ref{angezogenerop}). However, the
eigenvalues are real analytic functions of $\epsilon$ and we can label
the eigenvalues by the $l$-value they have when $\epsilon$ tends to
zero. In short, the only change needed in (\ref{obenpunch}) to
(\ref{obenpunch6}) is to replace $(2l+1)e^H_{n,l,\epsilon,Z}$ by the
sum of the eigenvalues in the multiplet of $H^H_{\epsilon,Z}$ that
converge to $e^H_{n,l,0,Z}$ as $\epsilon$ tends to zero. Since $W$ is
bounded by $1$, all our previous bounds for eigenvalue differences
(Lemmata \ref{l0} and \ref{l1}) continue to hold and we are finally
led to the $\liminf_{\epsilon\searrow}0$ in (\ref{obenpunch4}).

The crucial point is this: Even if $W$ is not spherical symmetric, the
sum of the eigenvalues in any multiplet is rotationally invariant to
first order in $\epsilon$ in the following sense. The only property of
$W$ that matters -- to first order -- is the average
$W_{average}:=(4\pi)^{-1} \int W(\omega)d\omega$.

Reversing the sign of $\epsilon$ again gives the lower bound.

2. {\it Proof of the weak convergence}: The proof can be rolled back
to the previous case as follows: First we assume that $V$ is spherically
symmetric and integrate the inequality
\begin{eqnarray*}
&&a^2\rho_Z(a\omega)\\
&\leq& \left[\sum_{l=0}^\infty q(2l+1)
\frac{\tr(H^H_{l,0,Z})_- - \tr(H^H_{l,\epsilon,Z})_-}{\epsilon Z^2}\theta(L
-l)\right.\\
&& \left.+\sum_{l=0}^\infty q(2l+1)
\frac{\tr(H_{l,0,Z})_- - \tr(H_{l,\epsilon,Z})_-}{\epsilon Z^2}\theta
(l-L)\right]
+ \const/(\epsilon Z^{\frac1{24}})
\end{eqnarray*}
against $aV(a)$ from $0$ to $\infty$. Observe that because of Lemma
\ref{l0} the summand of the sum on the right of side of this
integrated inequality is uniformly bounded by
$$ \frac9{(l+1)^2(2l+1)} \int_0^\infty  V(a) a^2 d a$$
which, when multiplied by $(2l+1)$, is summable. Again, the same argument
holds
when expressing the traces as sum over eigenvalues. Thus we are allowed to
take the limits term by term for the differences of the eigenvalues giving
the desired result as above.

The extension to the non-spherical case is as in Part 1.
\qed

\section{Extensions to Molecules \label{s4}}
The ground state energy of a neutral molecule with nuclear
charges $Z_1=\lambda z_1,...,Z_K=\lambda z_K$ and  positions of the nuclei at
${\frak R}_1,...,{\frak R}_K$ is given as
\begin{equation}
E(N,\vec Z) = \inf\{\inf\sigma(H_{N,\vec Z, \vec R})|\vec R\in
\rz^{3K}\}\label{ENZ}
\end{equation}
where
\begin{equation}
H_{N,\vec Z,\vec R} = \sum_{\nu=1}^N\left(-\Delta_\nu
- \sum_{\kappa=1}^K {Z_\kappa \over |{\frak r}_\nu-{\frak R}_\kappa|}\right)
+ \sum_{\mu,\nu=1 \atop \mu<\nu}^N{1 \over |{\frak r}_\mu-{\frak r}_\nu|}
+\sum_{\kappa,\kappa'=1 \atop \kappa<\kappa'}{Z_\kappa Z_{\kappa'}\over
|{\frak R}_\kappa-{\frak R}_{\kappa'}|}
\label{H}
\end{equation}
self-adjointly realized in ${\frak H}_N$. Here $\vec Z$ denotes the
$K$-tuple $(Z_1,...,Z_K)$ and $\vec R$ the $3K$-tuple
$(\mathfrak{R_1},...,\mathfrak{R}_K)$. We also set $\vec z
:=(z_1,...,z_K)$. Solovej \cite{Solovej1994} showed recently that for
arbitrary but fixed $\vec z$ and $N=Z_1+...+Z_K$
\begin{equation}
E(N,\vec Z) = \sum_{\kappa=1}^K E(Z_\kappa,Z_\kappa) +
o(\lambda^{\frac53})\label{solovej}
\end{equation}
holds as $\lambda$ tends to infinity and that the minimizing
inter-nuclear distances are of order $\lambda^{-5/21}$ or
bigger. These results imply among other things not only that the
atomic Scott correction and Schwinger correction implies the molecular
one but allows us to generalize Theorem \ref{t1} as well: The
molecular density in the vicinity of each nucleus converges in the
sense of Theorem 1 to the hydrogen density at each of the centers.
Our precise result is:
\begin{theorem}\label{t2}
Assume that $E(N,\vec Z)$ as defined in (\ref{ENZ}) is equal to
\begin{equation}
\inf\{\inf\sigma(H_{N,\vec Z, \vec R})| \vec R \in \rz^{3K},
\forall_{1\leq\kappa < \kappa'\leq K} |{\frak R}_\kappa - {\frak R}_\kappa'|
\geq R:= \const \lambda^\gamma\}\label{minimum}
\end{equation}
with $\gamma > -1/4$. Assume $N=Z_1+...+Z_k$, $Z_1=\lambda z_1,...,Z_K=\lambda
z_K$
with given fixed $z_1,...,z_K$. Furthermore fix $\kappa_0\in {1,...,K}$
and pick a sequence of ground state density matrices $d_c$ of $H_{N,\vec Z,
\vec R}$
with densities $\rho_{d_c}$. Define
$\rho_{\lambda,\kappa_0}({\frak r}) := \rho_\lambda(({\frak r}
- {\frak R}_{\kappa_0})/\lambda)/\lambda^3$. Finally assume
$W\in L^2({\mathbb S}^2)$. Then
\begin{equation}
\int_{{\mathbb S}^2} W(\omega) \rho_{\lambda,\kappa}(r\omega) \rightarrow
q \rho^H(r)\int_{{\mathbb S}^2} W.
\end{equation}
\end{theorem}
{\it Proof}: First note that by suitable relabeling we can always assume that
$\kappa_0=1$. Set $H_{N,\vec Z, \vec R}^\epsilon := H_{N,\vec Z,\vec R} -
\sum_{\nu=1}^N \epsilon U_\lambda({\frak r}-{\frak R}_1)$. Because of
(\ref{solovej}) it suffices that
$$\tr(d H_{N,\vec Z,\vec R}^\epsilon) \geq \tr (H_1-\epsilon U_\lambda)_+
\sum_{\kappa=1}^N \tr(H_\kappa)_- - D(\rho^{TF}_{Z_\kappa},
\rho^{TF}_{Z_\kappa})
- \const Z^{2-\delta}$$ for some positive $\delta$ and an approximate ground
state $d$. To this end let us introduce the localizing functions
$$\upsilon_\kappa({\frak r}) := \cos(\psi(|{\frak r}-{\frak R}_\kappa|/R))$$
where $\psi(t)$ is some continuous, piecewise differentiable, monotone
decreasing function which vanishes, if $t<\frac14$, and which is $\pi/2$,
if $t>\frac12$. Note that the supports of these functions have at
most finitely many points in common because $R$ is the minimal nuclear
distance. We also define
$$\upsilon_0 := \sqrt{1- \sum_{\kappa=1}^K \upsilon_\kappa^2}.$$

Now pick the density $\rho({\frak r}) := \sum_{\kappa=1}^K
\rho_{Z_\kappa}^{TF}(|{\frak r} - {\frak R}_\kappa|)$ and denote the
one-particle density matrix belonging to $d$ by $d_1$. -- Note that
$\tr d_1 = N$. -- By the correlation inequality
\cite{LiebThirring1975,Lieb1981} and
the localization formula using the above decomposition of unity we have
\begin{multline}
\tr(H_{N,\vec Z,\vec R}^\epsilon d)\\
\geq \tr\left\{\left[-\Delta_3-\sum_{\kappa'=1}^K\left(
\frac{Z_\kappa}{|.-{\frak R_\kappa}|} - \rho_\kappa\right) - \epsilon
U_Z\right]d_1\right\} -D(\rho,\rho)\\
+\sum_{\kappa,\kappa'=1\atop \kappa<\kappa'}^K{Z_\kappa Z_{\kappa'} \over
|{\frak R}_\kappa - {\frak R}_{\kappa'}|} -\const \lambda^{\frac53}\\
\geq \tr\left\{\sum_{\kappa=0}^K\upsilon_{\kappa}\left[-\Delta_3-
\sum_{\kappa'=1}^K\varphi^{TF}_{Z_{\kappa'}}(|.-{\frak R}_{\kappa'}S|)
- \epsilon U_Z\right]d_1 \upsilon_{\kappa'} \right\}\\
- \|\sum_{\kappa=0}^N |\grad\upsilon_\kappa|^2\|_\infty N
-\sum_{\kappa'=1}^K D(\rho_\kappa,\rho_{\kappa})-\const
\lambda^{\frac53}.\label{50}
\end{multline}
In (\ref{50}) we used the spherical symmetry of $\varphi_1,...,\varphi_K$ to
show
that $D(\rho_\kappa,\rho_{\kappa'}) \leq Z_\kappa Z_{\kappa'}|{\frak R}_\kappa
- {\frak R}_{\kappa'}|^{-1}$.

Now pick any arbitrary pair of different indices
$\kappa,\kappa'\in\{1,...,K\}$.
On the support of of $\upsilon_\kappa$ we have
$$\varphi^{TF}_{Z_{\kappa'}}(|{\frak r}-{\frak R}_{\kappa'}|) \leq
\frac{2^2 3^4\pi^2}{q^2(R/2)^4}$$
where we use the fact that the
Sommerfeld solution of the Thomas-Fermi equation is a pointwise upper
bound of the Thomas-Fermi potential (\cite{LiebSimon1977}, Section
V.2). Thus on the support of $\upsilon_\kappa$ we have
\begin{equation*}
\sum_{\kappa'=1, \kappa'\neq \kappa}^K \varphi^{TF}_{Z_\kappa}(|{\frak r}
- {\frak R}_\kappa|) \leq  {2^6 3^4 (K-1) \over R^4}.
\end{equation*}

For the derivative of the $\upsilon_\kappa$, which governs the localization
error,
we have the following uniform estimate: w$\sum_{\kappa=0}^K |\grad
\upsilon_{\kappa}|^2 = |\psi'(|{\frak r} - {\frak R}_\kappa|)|^2/R^2 =
4\pi^2$ where, for definiteness, we picked $\psi$ to be the linear
functions interpolating between $0$ and $\frac\pi2$ on the interval
$[\frac14,\frac12]$. Note that outside the annuli of thickness $R/2$
centered at the nuclei the derivatives vanish, in fact, whereas
in these annuli the bound is actually an equality.

This yields
\begin{multline}
\tr(H_{N,\vec Z,\vec R}^\epsilon d)\\
 \geq \sum_{\kappa=1}^K\left\{\tr\left[\left(-\Delta - \varphi^{TF}_{Z_\kappa}
- \frac{4\pi^2}{R^2}+\frac{2^4 3^4 \pi^2}{R^4}\chi_{B_{\frac R2}(0)}
\right)d_1\right] - D(\rho_\kappa,\rho_\kappa)\right\} -\const
\lambda^{\frac53}\\
\geq \sum_{\kappa=1}^K \left\{\tr\left(-\Delta -\varphi^{TF}_{Z_\kappa}
\right)_- - D(\rho_\kappa,\rho_\kappa)\right\}
- N\left(\frac{4\pi^2}{R^2}+\frac{2^4 3^4 \pi^2}{R^4}\right) -\const
\lambda^{\frac53}\\
\geq \inf\sigma(H_{Z_1,Z_1}^\epsilon)+\sum_{\kappa=2}^K
\inf\sigma(H_{Z_\kappa,Z_\kappa}
- \const \left({\lambda \over R^4} + \lambda^{5/3}\right)\\
\geq \inf\sigma(H_{Z_1,Z_1}^\epsilon)+\sum_{\kappa=2}^K
\inf\sigma(H_{Z_\kappa,Z_\kappa})
- o(\lambda^2).
\label{au}
\end{multline}
Combining this with Solovej's upper bound reduces the converge question to that
of the one-center case.\qed

\begin{appendix}
\section{Appendix: Facts about the Atomic Ground State Energy\label{appendix1}}
According to \cite{SiedentopWeikard1987O} we have
\begin{equation}
E_{Z,Z} \leq  E_{TF}(Z,Z) + \frac q8 Z^2 + \const Z^{\frac{47}{24}},
\label{Scottoben}\end{equation}
and according to \cite{SiedentopWeikard1989}
(see also \cite{SiedentopWeikard1991} and Hughes \cite{Hughes1990})
\begin{eqnarray}
E_{Z,Z} &\geq& \sum_{l=0}^{L-1}q(2l+1)\tr\left(H_{l,0,Z}^H\right)_- \nonumber
\\
&& +\sum_{l=L}^\infty q(2l+1)\tr\left(H_{l,0,Z}\right)_- -
D(\rho_{TF},\rho_{TF}) - \const Z^{\frac53}\nonumber \\
&\geq& E_{TF}(Z,Z) + \frac q8 Z^2 - \const Z^{\frac{17}9}\log
Z\label{Scottunten}
\end{eqnarray}
with $L=[Z^{\frac19}]$.
Combining (\ref{Scottoben}) and (\ref{Scottunten}) gives
\begin{eqnarray}
E_{Z,Z} &=& \sum_{l=0}^{L-1}q(2l+1)\tr\left(H_{l,0,Z}^H\right)_- \nonumber \\
&\phantom{=}& + \sum_{l=L}^\infty q(2l+1)\tr\left(H_{l,0,Z}\right)_- -
D(\rho_{TF},\rho_{TF}) + O(Z^{\frac{47}{24}}).\label{Scott}
\end{eqnarray}
\end{appendix}

\noindent
     Matematisk institutt,
        Universitetet i Oslo,
        Postboks 1053,
        N-0316 Oslo,
        Norway

\noindent
  Departments of Mathematics and Physics,
        Princeton University,
        Princeton, NJ 08544-0708,
        USA

\noindent
     Matematisk institutt,
        Universitetet i Oslo,
        Postboks 1053,
        N-0316 Oslo,
        Norway

\begin{thebibliography}{10}

\bibitem{Bach1989}
V. Bach.
\newblock A proof of {S}cott's conjecture for ions.
\newblock {\em Rep. Math. Phys.}, 28(2):213--248, October 1989.

\bibitem{FeffermanSeco1992}
C.~Fefferman and L.~Seco.
\newblock Eigenfunctions and eigenvalues of ordinary differential operators.
\newblock {\em Adv. Math.}, 95(2):145--305, October 1992.

\bibitem{FeffermanSeco1994T}
C.~Fefferman and L.~Seco.
\newblock The density of a one-dimensional potential.
\newblock {\em Adv. Math.}, 107(2):187--364, September 1994.

\bibitem{FeffermanSeco1994Th}
C.~Fefferman and L.~Seco.
\newblock The eigenvalue sum of a one-dimensional potential.
\newblock {\em Adv. Math.}, 108(2):263--335, October 1994.

\bibitem{FeffermanSeco1994}
C.~Fefferman and L.~Seco.
\newblock On the {D}irac and {S}chwinger corrections to the ground-state energy
  of an atom.
\newblock {\em Adv. Math.}, 107(1):1--188, August 1994.

\bibitem{FeffermanSeco1995}
C.~Fefferman and L.~Seco.
\newblock The density in a three-dimensional radial potential.
\newblock {\em Adv. Math.}, To appear, 1995.

\bibitem{FeffermanSeco1993}
C.~L. Fefferman and L.~A. Seco.
\newblock Aperiodicity of the {H}amiltonian flow in the {T}homas-{F}ermi
  potential.
\newblock {\em Revista Mathem\'atica Iberoamericana}, 9(3):409--551, 1993.

\bibitem{HeilmannLieb1995}
O.~J. Heilmann and E.~H. Lieb.
\newblock The electron density near the nucleus of a large atom.
\newblock {\em Preprint}, 1995.

\bibitem{Hughes1990}
W. Hughes.
\newblock An atomic lower bound that agrees with {S}cott's correction.
\newblock {\em Adv. in Math.}, 79:213--270, 1990.

\bibitem{IvriiSigal1993}
V.~J. Ivrii and I.~M. Sigal.
\newblock Asymptotics of the ground state energies of large {C}oulomb systems.
\newblock {\em Annals of Math.}, 138(2):243--335, 1993.

\bibitem{Lieb1981}
E.~H. Lieb.
\newblock {T}homas-{F}ermi and related theories of atoms and molecules.
\newblock {\em Rev. Mod. Phys.}, 53(4):603--641, October 1981.

\bibitem{LiebSimon1977}
E.~H. Lieb and B. Simon.
\newblock The {T}homas-{F}ermi theory of atoms, molecules and solids.
\newblock {\em Adv. Math.}, 23:22--116, 1977.

\bibitem{LiebThirring1975}
E.~H. Lieb and W.~E. Thirring.
\newblock Bound for the kinetic energy of {F}ermions which proves the stability
  of matter.
\newblock {\em Phys. Rev. Lett.}, 35(11):687--689, September 1975.

\bibitem{Olver1961}
F.~W.~J. Olver.
\newblock Error bounds for the {L}ioville-{G}reen (or {WKB}) approximation.
\newblock {\em Proc. Camb. Phil. Soc.}, 57:790--810, 1961.

\bibitem{Olver1962}
F.~W.~J. Olver.
\newblock {\em Tables for Bessel Functions of Moderate or Large Orders},
  volume~6 of {\em Mathematical Tables}.
\newblock Her Majesty's Stationary Office, London, 1 edition, 1962.

\bibitem{Olver1968}
F.~W.~J. Olver.
\newblock Bessel functions of integer order.
\newblock In Milton Abramowitz and Irene~A. Stegun, editors, {\em Hanbook of
  Mathematical Functions with Formulas, Graphs, and Mathematical Tables},
  chapter~9, pages 355--433. Dover Publications, New York, 5 edition, 1968.

\bibitem{Olver1974}
F.~W.~J. Olver.
\newblock {\em Asymptotics and Special Functions}.
\newblock Academic Press, New York, 1 edition, 1974.

\bibitem{ReedSimon1978}
M. Reed and B. Simon.
\newblock {\em Methods of Modern Mathematical Physics}, volume 4: Analysis of
  Operators.
\newblock Academic Press, New York, 1 edition, 1978.

\bibitem{Scott1952}
J.~M.~C. Scott.
\newblock The binding energy of the {T}homas-{F}ermi atom.
\newblock {\em Phil. Mag.}, 43:859--867, 1952.

\bibitem{Siedentop1994A}
H. Siedentop.
\newblock An upper bound for the atomic ground state density at the nucleus.
\newblock {\em Lett. Math. Phys.}, 32(3):221--229, November 1994.

\bibitem{Siedentop1994B}
H. Siedentop.
\newblock Bound for the atomic ground state density at the nucleus.
\newblock In {\em Proceedings of the Vancouver Summer School on Mathematical
  Quantum Mechanics, July 1993}, Providence, Rhode Island, To appear. American
  Mathematical Society.

\bibitem{SiedentopWeikard1987O}
H. Siedentop and R. Weikard.
\newblock On the leading energy correction for the statistical model of the
  atom: Interacting case.
\newblock {\em Commun. Math. Phys.}, 112:471--490, 1987.

\bibitem{SiedentopWeikard1989}
H. Siedentop and R. Weikard.
\newblock On the leading correction of the {T}homas-{F}ermi model: Lower bound
  -- with an appendix by {A.} {M.} {K.} {M\"u}ller.
\newblock {\em Invent. Math.}, 97:159--193, 1989.

\bibitem{SiedentopWeikard1991}
H. Siedentop and R. Weikard.
\newblock A new phase space localization technique with application to the sum
  of negative eigenvalues of {Schr\"odinger} operators.
\newblock {\em Annales Scientifiques de l'\'Ecole Normale Sup{\'e}rieure},
  24(2):215--225, 1991.

\bibitem{Solovej1994}
Jan~Philip Solovej.
\newblock In preparation.

\end{thebibliography}
\end{document}